\documentclass[twocolumn,showpacs,preprintnumbers,prl]{revtex4}
\usepackage{graphicx,epsfig}

\def\simgt{\rlap{\lower 3.5 pt \hbox{$\mathchar \sim$}} \raise 1pt \hbox {$>$}}
\def\simlt{\rlap{\lower 3.5 pt \hbox{$\mathchar \sim$}} \raise 1pt \hbox {$<$}}

\begin{document}

\preprint{PITHA 05/19
}

\title{
\boldmath
Enhanced electroweak penguin amplitude in $B\to VV$ decays
\unboldmath}
\author{M.~Beneke${}^1$,  J.~Rohrer${}^1$ and D.~Yang${}^2$}

\affiliation{
$^1\!\!\!$ Institut f\"ur Theoretische Physik E, RWTH Aachen,
D-52056 Aachen, Germany\\
$^2\!\!\!$ Department of Physics, Nagoya University,
Nagoya 464-8602, Japan
}

\date{December 19, 2005}

\begin{abstract}
\noindent
We discuss a novel electromagnetic penguin contribution to the 
transverse helicity amplitudes in $B$ decays to two vector mesons, 
which is enhanced by two powers of $m_B/\Lambda$ relative 
to the standard penguin amplitudes. This leads to unique polarization 
signatures in penguin-dominated decay modes such as $B\to\rho K^*$ 
similar to polarization effects in the radiative 
decay  $B\to K^*\gamma$, and offers new opportunities to probe 
the magnitude and chirality of flavour-changing neutral current 
couplings to photons.
\end{abstract}

\maketitle

\section{Introduction}

\noindent
Decays of $B$ mesons into two charmless mesons provide an abundant 
source of information on flavour- and CP-violating phenomena in the weak 
interactions of quarks. In particular, decays to two vector 
mesons ($B\to VV$) can shed light on the helicity structure of these
interactions through polarization studies. While predicted to be 
fundamentally V-A in the Standard Model (SM), a deviation from this 
expectation cannot currently be excluded. The first observations 
of $B\to VV$ decays show no anomalies in the helicity structure, 
but point to a reduced amount of longitudinal polarization in 
penguin-dominated decays~\cite{Aubert:2003mm}. 
This has led to theoretical studies 
that reconsider strong interactions effects in $B\to VV$ 
decays~\cite{Kagan:2004uw,Colangelo:2004rd,bry}, or invoke new fundamental 
interactions~\cite{np}.

Any particular $B\to VV$ decay is characterized by the three helicity 
amplitudes $A_0$ (longitudinal), $A_-$, and $A_+$. A quark model or naive 
factorization analysis~\cite{Korner:1979ci} 
leads to the expectation that for $\bar B$, 
i.e. $b$-quark, decay the helicity amplitudes are in proportions 
\begin{equation}
A_0:A_-:A_+ = 1 : \frac{\Lambda}{m_b} : 
\left(\frac{\Lambda}{m_b}\right)^{\!2}
\label{hierarchy}
\end{equation}
with $\Lambda\approx 0.5\,$GeV the strong interaction scale and 
$m_b\approx 5\,$GeV the bottom quark mass. This expectation has been  
parametrically (not necessarily numerically)
confirmed~\cite{Kagan:2004uw} in the framework of QCD factorization, 
which provides a theoretical basis for the heavy-quark expansion 
of $B$ decays to charmless mesons~\cite{BBNS1}. The hierarchy 
(\ref{hierarchy}) of helicity amplitudes follows from the V-A 
structure of the standard weak interactions.

In this Letter we point out and discuss an effect which has been 
neglected in all previous studies of $B\to VV$, but which substantially 
alters the prediction for polarization observables. 
The effect is connected with electromagnetic penguin transitions, 
and appears only for neutral vector mesons. It leads to the unique 
feature that the transverse electroweak penguin amplitude is dominated
by the electromagnetic dipole operator providing a signature similar to
polarization in radiative decays $B\to
K^\ast\gamma$~\cite{Mannel:1997pc}, 
but which is easier to access experimentally. 

\begin{figure}[b]
   \vspace{-2.4cm}
   \hspace*{-5.3cm}
   \epsfxsize=20cm
   \epsffile{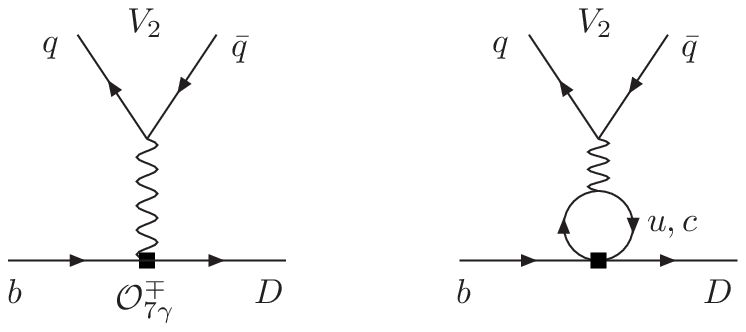}
   \vspace*{-23.4cm}
\caption{\label{fig1} 
Leading contributions to $\Delta \alpha_{3,\rm EW}^{p\mp}(V_1 V_2)$ 
defined in the text.}
\end{figure}
The effect in question is related to the two diagrams shown in 
Figure~\ref{fig1}. When the vector meson $V_2$ is transversely 
polarized, there exists a large contribution to the decay 
amplitude due to the small virtuality $m_{V_2}^2$ of the 
intermediate photon propagator. This is in contrast to the case 
of longitudinal polarization, where the photon propagator is 
canceled, and the amplitude is local on the scale 
$m_b$~\cite{Beneke:2003zv}. The large transverse amplitude is 
best described by a short-distance transition $b\to D\gamma$ 
($D=d,s$), followed by the transition of the low-virtuality photon 
($q^2\ll m_b^2$) to the neutral vector meson. We shall perform a 
factorization analysis of the amplitude below. 

The calculation of the diagrams in Figure~\ref{fig1} is 
straightforward. The weak interactions are given in terms of 
the standard effective Hamiltonian~\cite{Buchalla:1995vs}. 
We use the conventions of~\cite{Beneke:2001ev}, but generalize 
the electromagnetic dipole operators to include both 
chiralities 
\begin{eqnarray}
{\cal H}_{\rm eff} &=& \frac{G_F}{\sqrt{2}} 
\sum_{p=u,c} \lambda_p^{(D)} \sum_{a=-,+} C_{7\gamma}^a Q_{7\gamma}^a  
+\ldots,\\
Q_{7\gamma}^\mp &=&
-\frac{e\bar{m}_b}{8 \pi^2} \,\bar
D\sigma_{\mu\nu}(1\pm \gamma_5)F^{\mu\nu} b,
\label{o7}
\end{eqnarray}
where $\lambda_p^{(D)}=V_{pb}V_{pD}^\ast$. 
The ellipses denote other operators (see~\cite{Beneke:2001ev}). 
In the SM $C^+_{7\gamma}$ 
is suppressed by a factor $m_D/m_b$, hence $Q_{7\gamma}^+$ is 
usually neglected. The remaining term is then simply denoted by 
$C_{7\gamma} Q_{7\gamma}$. However, in generic 
extensions of the SM, there is no reason to expect 
a suppression of additional contributions to $C^+_{7\gamma}$ 
relative to  $C^-_{7\gamma}$. The coupling of the photon to the 
quark electric charge in $V_2$ implies that the diagrams of 
Figure~\ref{fig1} contribute to the electroweak penguin amplitude 
in the general flavour decomposition of hadronic two-body decay 
amplitudes. Adopting the $\alpha_i$ notation of~\cite{Beneke:2003zv} 
extended to allow for the three helicity amplitudes of $B\to VV$, the 
new contribution to the transverse electroweak penguin amplitudes 
is 
\begin{equation}
\Delta \alpha_{3,\rm EW}^{p\mp}(V_1 V_2) =  \mp
\frac{2\alpha_{\rm em}}{3\pi}\,C_{7\gamma,\rm eff}^\mp \,R_\mp\,
\frac{m_B\bar{m}_b}{m_{V_2}^2}
\label{dal3ew}
\end{equation}
with $C_{7\gamma,\rm eff}^\mp$ taking into account the effect of quark 
loop diagrams (see Figure~\ref{fig1}). $R_\mp$ is a ratio of tensor 
to (axial) vector $B\to V_1$ form factors such that $R_-$ equals 1 in 
the heavy quark limit~\cite{Charles:1998dr}, while $R_+$ is of 
order $m_b/\Lambda$. We note the large enhancement factor
$m_B\bar{m}_b/m_{V_2}^2  
\sim (m_b/\Lambda)^2$, which implies that the first hierarchy 
in (\ref{hierarchy}) is inverted, rendering the negative-helicity 
amplitude $A_-$ leading over the longitudinal amplitude $A_0$ in the
heavy-quark limit. Of course, for real values of $m_b/m_{V_2}$ this 
enhancement is compensated by the small electromagnetic coupling 
$\alpha_{\rm em} =e^2/(4\pi)$. For instance, for neutral $\rho$
mesons, we obtain 
$\Delta \alpha_{3,\rm EW}^{p-}(K^*\rho)\approx 0.02$. 
This should be compared to the uncorrected negative-helicity electroweak 
penguin amplitude 
\begin{equation}
\alpha_{3,\rm EW}^{p-}(K^*\rho) = 
C_7+C_9+\frac{C_8+C_{10}}{N_c}+\ldots 
\approx -0.01,
\end{equation}
and the leading QCD penguin amplitude 
\begin{equation}
\hat\alpha_4^{c-}(\rho K^*) = C_4+\frac{C_3}{N_c} +\ldots 
\approx -0.055.
\end{equation}
The $C_i$ are Wilson coefficients for the various penguin 
operators in the effective Hamiltonian~\cite{Buchalla:1995vs}, and 
the ellipses denote the 1-loop corrections in QCD factorization 
\cite{bry}, which we have taken into account in the numerical
estimates. In the SM the corresponding positive-helicity 
amplitudes are suppressed by about an order of magnitude 
relative to the negative-helicity ones as explained above. 

There are strong-interaction corrections to the leading-order 
expression (\ref{dal3ew}) from gluon exchange between the 
quark lines in the second diagram of Figure~\ref{fig1}, and also 
through hard interactions with the spectator quark (not shown 
in the Figure) in the $B$ meson. Due to factorization as discussed below, 
these corrections modify only the effective $b\to D\gamma$ 
transition at leading order in the expansion in $\Lambda/m_b$. 
They have been computed in next-to-leading order in the context 
of factorization of exclusive radiative $B$
decays~\cite{Beneke:2001at}, and can be incorporated 
by substituting $C_{7\gamma}^-\to {\cal C}_7^\prime$ 
(first paper of~\cite{Beneke:2001at}, eq.~(62)). 
Turning this argument around, 
the absolute value of $\Delta \alpha_{3,\rm EW}^{c-}(K^* V_2)$ 
can be obtained from the branching fraction of $B\to K^\ast\gamma$ 
via 
\begin{eqnarray}
&&\left|\Delta \alpha_{3,\rm EW}^{c-}(K^\ast V_2)\right| =  
\frac{2\alpha_{\rm em}}{3\pi}\,R_-\,\frac{m_B^2}{m_{V_2}^2}
\nonumber\\
&&\hspace*{1cm}\times\left(\frac{\Gamma(B\to K^\ast\gamma)}{\displaystyle
\frac{G_F^2|V^*_{ts} V_{tb}|^2}{8\pi^3} \,
\frac{\alpha_{\rm em}}{4\pi}\,m_B^5\,T_1^{K^\ast}\!(0)^2}\right)^{\!1/2}
\end{eqnarray}
with $T_1^{K^\ast}(0)\approx 0.28$ a tensor form factor. This 
results in $|\Delta \alpha_{3,\rm EW}^{c-}(K^\ast \rho)| 
=0.023$, close to the leading-order estimate from (\ref{dal3ew}).

We therefore conclude that the new radiative contribution to the 
negative-helicity electroweak penguin amplitude is at least twice 
as large (and opposite in sign) as was previously assumed. 
For penguin-dominated $b\to s$ transitions it is almost half the size 
of the leading QCD penguin amplitude, and should therefore have 
visible impact on polarization measurements. In case of new 
interactions generating $C_{7\gamma}^+$, the corresponding 
contribution to the positive-helicity amplitude (\ref{dal3ew}) 
should be observed against a very small Standard Model background. 
 
\section{Factorization analysis}

\noindent
Since the existence of an amplitude violating the power 
counting~(\ref{hierarchy}) may appear surprising, we sketch 
how this amplitude emerges and factorizes in soft-collinear 
effective theory (SCET)~\cite{Bauer:2000yr}. The notation and 
method of the following discussion is similar to the one 
in~\cite{Beneke:2003pa}. After integrating out the scale 
$m_b$, SCET formalizes the interaction of the static $b$-quark 
field $h_v$ with collinear fields for the light-like 
direction $n_-$, in which meson $V_1$ moves, and collinear 
fields for the light-like direction $n_+$ of meson $V_2$. 
Let $\chi$ denote the collinear quark field corresponding 
to $V_2$, and let $V_2$ be the meson that does not pick up the 
spectator quark from the $B$ meson. The leading quark bilinears 
that have non-vanishing overlap with $\langle V_2|$ are 
\begin{equation}
\bar\chi \!\not\!n_- (1\mp \gamma_5)\chi,\quad
\bar\chi \!\not\!n_- \gamma_\perp^\mu (1\pm \gamma_5)\chi.
\end{equation}
The subscript $\perp$ denotes projection of a Lorentz vector on 
the plane transverse to the two light-cone vectors $n_\mp$. 
Both operators scale as $\lambda^4$ according to the SCET scaling 
rules; the first overlaps only with the longitudinal polarization 
state of $V_2$, the second only with a transverse vector meson. 
However, the second operator is not generated by the V-A 
interactions of the SM (at least at the tree and 1-loop level). 
This implies the power suppression of $A_\mp$ relative to 
$A_0$ in (\ref{hierarchy}), since the leading contribution to 
transverse polarization now involves an operator with an 
additional derivative $D_\perp\sim \lambda^2\sim \Lambda/m_b$.
 
This reasoning ignores electromagnetic effects. Including QED 
in SCET, there is a collinear photon field with unsuppressed 
interactions with collinear quarks (of the same direction). Only 
the transverse photon field is truly a degree of freedom of the 
theory, since the other two components are either gauge-artefacts, or
can be eliminated by the field equations. Hence there is an 
additional operator $e A^\mu_{\gamma_\perp}=W_\gamma^\dagger 
iD_{\gamma\perp}^\mu W_\gamma$ (where $W_\gamma$ is an electromagnetic 
Wilson line formally required to make the operator gauge-invariant), 
which overlaps only with a transversely polarized vector meson. 
To first order in the electromagnetic coupling the matrix 
element can be computed exactly yielding
\begin{equation}
\langle V_2|[W_\gamma^\dagger iD_{\gamma\perp}^\mu W_\gamma](0)|0\rangle = 
-\frac{2i}{3} \,a_{V_2} \frac{e^2 f_{V_2}}{m_{V_2}}\,
\epsilon_\perp^{\ast\mu}
\label{me2}
\end{equation}
with $\epsilon_\perp^\mu$ a transverse polarization vector, $f_{V_2}$ 
the decay constant, and $a_{V_2}$ a constant that depends on the 
quark-flavour composition of $V_2$, $a_\rho=3/2$, $a_\omega=1/2$, 
$a_\phi=-1/2$. (The convention for the covariant derivative
corresponding to (\ref{o7}) is $iD_{\gamma}^\mu = 
i\partial^\mu+e_q A^\mu_\gamma$ with $e_q$ the quark electric 
charge.) The crucial 
point is that the operator $W_\gamma^\dagger 
iD_{\gamma\perp}^\mu W_\gamma$ scales with 
$\lambda^2$, hence this contribution to $A_\mp$ is a factor 
$m_b/\Lambda$ larger than even the longitudinal amplitude $A_0$.
Thus, we find the tree-level matching equation (see also 
\cite{Becher:2005fg})
\begin{equation}
Q_{7\gamma}^\mp \to -\frac{\bar m_b m_B}{4\pi^2} \,
\big[\bar\xi W \gamma_{\perp\mu} (1\mp \gamma_5)h_v\big](0)
\big[W_\gamma^\dagger 
iD_{\gamma\perp}^\mu W_\gamma\big](0),
\label{scetop} 
\end{equation}
valid as an equation for the $\langle V_1 V_2|\ldots|\bar B\rangle$ 
matrix element. In SCET only soft fields can couple to the two
brackets representing collinear field products in the two different 
directions. But since the photon operator in the second bracket 
is a colour-singlet, the soft fields decouple, and the matrix element 
of the right-hand side of (\ref{scetop}) falls apart into 
(\ref{me2}) and $\langle V_1|
\bar\xi \,W \gamma_\perp^\mu (1\mp \gamma_5)h_v|\bar B\rangle$, 
which is proportional to the SCET form factor 
$\xi_\perp$~\cite{Charles:1998dr} at maximal recoil. 
Eq.~(\ref{scetop}) has to be amended by radiative corrections 
as well as a second operator structure with an additional 
transverse derivative in the first bracket. This is very similar 
to heavy-to-light form factors~\cite{Beneke:2003pa}, in fact, 
these corrections simply restore the QCD tensor form factor. 
Combining (\ref{me2},\ref{scetop}), we therefore find 
\begin{eqnarray}
&&\langle V_1 V_2|C_{7\gamma}^\mp \,Q_{7\gamma}^\mp|\bar B\rangle
= i m_{V_2} m_B 2 T_1^{V_1}(0) f_{V_2} a_{V_2} 
\nonumber\\
&&\hspace*{1cm}\times \left(\mp
\frac{2\alpha_{\rm em}}{3\pi}\right) C_{7\gamma}^\mp \,
\frac{m_B\bar{m}_b}{m_{V_2}^2},
\label{new2}
\end{eqnarray}
which on accounting for the normalization of $\alpha_{3,\rm EW}^{p,h}$ 
\cite{bry,Beneke:2003zv} reproduces (\ref{dal3ew}). The previous
equation should be understood such that the matrix element of 
$Q_{7\gamma}^-$ ($Q_{7\gamma}^+$) takes the value given only when 
both $V_1$ and $V_2$ have negative (positive) helicity, but 
is zero otherwise. In general, the four-quark operators from the 
effective weak Hamiltonian also contribute to the matching coefficient
of the SCET operator on the right-hand side of (\ref{scetop}), 
and including further spectator-scattering effects replaces 
$C_{7\gamma}^-$ by ${\cal C}_7^\prime$ as discussed above.

\section{\boldmath The $B\to \rho K^\ast$ system}

\noindent
We now focus on the eight $B\to \rho K^*$ decay modes, where the 
electroweak penguin amplitude is largest relative to the leading 
QCD penguin amplitude ($a_\rho=3/2$). Assuming isospin symmetry, 
the $\rho K^*$ system is described by six complex strong interaction 
parameters for each helicity $h=0,-,+$. Neglecting the 
colour-suppressed electroweak penguin 
amplitude and the doubly CKM suppressed QCD penguin amplitude is 
a good approximation for elucidating the effect of the new 
(colour-allowed) electroweak penguin contribution, hence we 
write 
\begin{eqnarray}
A_h(\rho^- \bar K^{\ast 0}) &=& P_h
\nonumber\\
\sqrt{2}\,A_h(\rho^0 K^{\ast -}) &=& [P_h+P_h^{EW}]+e^{-i\gamma} \,[T_h+C_h]
\nonumber\\
A_h(\rho^+ K^{\ast -}) &=& P_h+e^{-i\gamma} \,T_h
\nonumber\\
-\sqrt{2}\,A_h(\rho^0 \bar K^{\ast 0}) &=&
[P_h-P_h^{EW}]+e^{-i\gamma}\,[-C_h],
\label{first}
\end{eqnarray}
and define $x_h=X_h/P_h$, where $P_h$ is the QCD penguin amplitude. 
The tree amplitudes $T_h$, $C_h$ are suppressed by the 
CKM factor $\epsilon_{\rm KM} = |V_{ub} V_{us}^*|/
|{V_{cb} V_{cs}^*}| \sim 0.025$. Assuming $\gamma=70^\circ$ is known, one 
can obtain $P_h$ from an angular analysis of the $\rho^- \bar
K^{\ast0}$ final state, $t_h$ from $\rho^\pm K^{\ast\mp}$, 
and $p_h^{EW}$ and $c_h$ from the remaining four decay modes. 
In principle, this allows for a determination of $P_h^{EW}$, 
which can be compared to the theoretical result. In practice, 
a complete amplitude analysis will be experimentally difficult.

The sensitivity to the electroweak penguin amplitude is made apparent 
in CP-averaged helicity-decay rate ratios such as 
\begin{equation}
S_h\equiv \frac{2 \bar \Gamma_h(\rho^0\bar K^{\ast 0})}
{\bar \Gamma_h(\rho^-\bar K^{\ast 0})} =
\left|1-p_h^{EW}\right|^2+\Delta_h,
\label{r1}
\end{equation}
where $\Delta_h$ depends on $c_h$ (and mildly on $p_h^{EW}$), and 
vanishes for $c_h\to 0$. To estimate $S_-$, we assume that 
the positive-helicity amplitudes are negligible as predicted in 
the SM, and use the observed $\rho^- \bar K^{\ast 0}$ branching 
fraction and longitudinal polarization fraction $f_L$ to determine 
the magnitude of $P_0$ and $P_-$. We shall also assume that the 
phase of $p^{EW}_h$ is not more than $30^\circ$ away from 
$0$ or $\pi$. Writing $p^{EW}_h=[P^{EW}_h/T_h]\times t_h$, this
amounts to the assumption that no large CP asymmetries will be 
found in $B\to \rho^\pm K^{*\mp}$. 
For all other quantities we perform 
a calculation in the QCD factorization framework. In this procedure 
there is a considerable uncertainty in $P_-$ due to the discrepant 
experimental results on $f_L(\rho^+ K^{*0})$ \cite{Aubert:2003mm}, 
which may result in an over-estimate of $P_-$ and hence an 
under-estimate of $p_-^{EW}$. It is therefore not excluded that 
the electromagnetic penguin effect is more pronounced than in 
the following theoretical estimates. Keeping this in mind, we find 
$\mbox{Re}\,(p^{EW}_-)={-0.23}\pm 0.08\,\,[{+0.14}^{+0.04}_{-0.05}]$ 
and $\Delta_-={-0.0}\pm 0.2$, 
yielding 
\begin{equation}
S_- ={1.5}\pm 0.2\,\,[{0.7}\pm 0.1].
\label{s1}
\end{equation} 
Here (and below) the numbers in brackets refer to the calculation 
without the new electromagnetic penguin contribution. Despite 
the current large theoretical uncertainties, which could be removed 
with more experimental data, eq.~(\ref{s1}) clearly 
shows the impact of this contribution on polarization observables. The 
effect is even more significant for the 
ratio of the two final states with neutral $\rho$ mesons, as 
$S_-/S_-^\prime$ [(\ref{r2}) below] changes by a factor of about 4 
whether or not the electromagnetic penguin contribution is included, 
but for this ratio the tree contamination is also larger. 
Data is currently not available to test (\ref{s1}), but
we may instead consider 
\begin{equation}
S_h^\prime\equiv \frac{2 \bar\Gamma_h(\rho^0\bar K^{\ast -})}
{\bar \Gamma_h(\rho^-\bar K^{\ast 0})} =
\left|1+p_h^{EW}\right|^2+\Delta_h^\prime.
\label{r2}
\end{equation}
Following the same strategy as above, we obtain 
$\Delta_-^\prime={-0.1}\pm 0.0$, and 
$S_-^\prime ={0.5}\pm 0.1\,\,[{1.2}\pm 0.1]$.
In the absence of direct CP asymmetries $S_h^\prime$ is 
directly related to the corresponding ratio of 
polarization fractions $f_h^\prime \equiv f_h(\rho^0\bar K^{\ast -})/
f_h(\rho^-\bar K^{\ast 0})$. Including a theoretical 
estimate of the CP asymmetries, we obtain 
\begin{eqnarray}
&&f_0^\prime = {1.3}\pm 0.1\,\,[{1.1}\pm 0.1], \\
&&f_-^\prime = \frac{1-f_L(\rho^0\bar K^{\ast -})}
{1-f_L(\rho^-\bar K^{\ast 0})} = 
0.4 \pm 0.1\,\,[{0.8}\pm 0.1]. \quad
\end{eqnarray}
This can be compared to the experimental values 
$f_0^\prime|_{\rm exp} = {1.45}^{+0.64}_{-0.58}$, 
$f_-^\prime|_{\rm exp} = {0.12}^{+0.44}_{-0.11}$~\cite{Aubert:2003mm}. 

Finally we comment on the possibility of detecting the presence 
of new flavour-changing neutral currents in the form of an 
electromagnetic penguin operator with opposite chirality,
$Q^+_{7\gamma}$. For this analysis, one must isolate experimentally 
the positive-helicity amplitudes. Theoretically, 
all positive-helicity amplitudes are suppressed, 
except for the electromagnetic penguin contribution 
$\Delta P_+^{EW}$ to the electroweak penguin amplitude. In the naive 
factorization approximation $X_+ =  r X_-$, where $r$ is a 
$\Lambda/m_b$-suppressed form factor ratio, while 
$\Delta P_+^{EW} \approx C^+_{7\gamma}/C^-_{7\gamma}\, 
\Delta P_-^{EW}$ is suppressed only by the ratio of Wilson 
coefficients (see (\ref{new2})). A conservative analysis of the 
$b\to s\gamma$ branching fraction constrains 
$C_{7\gamma}^+/C_{7\gamma}^-<0.5$, hence it is possible that 
the suppression is weak. This would lead to 
$P_+^{EW}\gg P_+$, in which case the positive-helicity 
decay rates of the $\rho^0 K^\ast$ final states are 
much larger than the 
$\rho^\pm K^*$ ones. A complete angular analysis of the $\rho K^\ast$ 
system should allow a determination of $p_+^{EW}$ even when 
it is not dominant, possibly allowing a limit on  
$C^+_{7\gamma}/C^-_{7\gamma}$ of order $r\approx 0.1$. 

In conclusion, we discussed an electromagnetic penguin contribution 
to non-leptonic $B$ decays that has previously been overlooked. 
It is the largest contribution to the negative-helicity
electroweak penguin amplitude, and substantially modifies the 
theoretical expectations for polarization observables in 
$b\to s$ penguin-dominated decays, in particular to the 
$\rho^0 K^\ast$ final states. These observables may therefore 
be of considerable interest to the search for electromagnetic 
flavour-changing neutral currents with chirality equal or opposite to 
the SM.


\begin{acknowledgments}
\noindent
We thank G.~Buchalla and M.~Neubert for comments. 
The work of M.B. is supported by the DFG SFB/TR~9 
``Computergest\"utzte Theo\-re\-ti\-sche Teilchenphysik''; 
the work of J.R. by GIF under Grant No. I - 781-55.14/2003.
D.Y. acknowledges support from the Japan Society for the 
Promotion of Science. 
\end{acknowledgments}

\end{document}